# Sub-nanometer measuring ellipticity of a suspended optical nanowaveguides based on nondegenerate mechanical modes

CHENXI WANG[1], LIJUN SONG[1], JIANTING WANG[1], JING ZHOU[1], KANGJIE FENG[1], QIANG ZHANG[1,2], CHANG LING ZOU[3,1], GANG LI[1,2], PENGFEI ZHANG[1,2,\*], AND TIANCAI ZHANG[1,2,\*\*]

[1]*State Key Laboratory of Quantum Optics and Quantum Optics Devices, Institute of Opto-electronics, Shanxi University, Taiyuan, Shanxi 030006, People's Republic of China*
[2]*Collaborative Innovation Center of Extreme Optics, Shanxi University, Taiyuan, Shanxi 030006, People's Republic of China*
[3]*Key Laboratory of Quantum Information, Chinese Academy of Sciences, University of Science and Technology of China, Hefei 230026, People's Republic of China*
\**zhangpengfei@sxu.edu.cn*
\*\**tczhang@sxu.edu.cn*

**Abstract:**  Optical waveguides with miniature dimensions to the nanoscale can facilitate the development of highly integrated photonic devices, integrated optical circuits and hybrid quantum system coupling with emitters. Nondegenerate intrinsic flexural mechanical modes of nanowaveguides provide unique insights into the mechanical properties and structural integrity of materials, which is great significance to the applications of the nanowaveguides. Here,we propose and implement a scheme to measure the nondegenerate intrinsic flexural mechanical modes of a suspended optical nanowaveguide, a tapered optical fiber (TOF). A TOF with an elliptical cross section can support two nondegenerate intrinsic flexural mechanical modes (IFMMs) because the two orthogonal modes vibrate along the principal axes (major or minor axis) of the elliptical TOF cross section with splitting vibration frequencies. The frequency ratio for the two IFMMs approaches a constant with increasing mode order, which is equal to the inverse of the TOF ellipticity. Thus, the TOF ellipticity can be determined on the basis of the splitting vibration frequencies of the nondegenerate modes with subnanometer-level accuracy, 0.16 nm for a TOF radius of 260 ± 5 nm. The elliptical TOF's nondegenerate IFMMs offer a novel pathway for research on nanoscale structures and vector measurement in fields such as quantum optics, atom physics, sensing, optical communications, and micro/nanomechanics.

## 1. Introduction

Optical waveguides with miniature dimensions to the nanoscale, called nanowaveguides, enable the guidance and manipulation of light in extremely confined spaces, facilitating the development of highly integrated photonic devices and integrated optical circuits. Through meticulous design of nanowaveguide structures, it is possible to control propagation optical modes effectively, which is crucial for beam shaping and the modulation of optical fields [1,2]. Minuscule cross-sections of nanowaveguides are harnessed to enhance light–matter interactions for investigating nonlinear optical effects, optical sensing and quantum phenomena [3–7]. Nanowaveguides play a key role in fundamental research and applied fields [8,9].

Tapered optical nanofibers (TOFs), as significant typical one-dimensional optical nanowaveguides, present extraordinary optical and mechanical properties due to their characteristic shapes and ultrahigh optical transmissivity [4, 10]. TOFs can be derived from conventional optical fibers through the flame-brushing technique [11]. TOFs have versatility and exceptional optical and mechanical properties, which facilitate advancements and innovations across various applications [7, 12–14], including near-field optics [15], nonlinear optics [16], atom optics [17], and optomechanics [18], as well as in the design of microcouplers [19], resonators [20] and sensors [21].

TOFs can support strongly confined optical evanescent fields and mechanical modes [18,22,23]. Under ideal conditions, optical or mechanical modes within TOFs with circular cross-sections naturally exhibit two degenerate modes for two orthogonal optical polarizations or mechanical vibration along the TOF principal axes. The two degenerate modes have the same properties. However, this symmetrical harmony can be disrupted easily in practical situations because of the breaking of the profile symmetry of TOFs. This results in nondegenerate intrinsic modes. These factors include inherent defects introduced from the manufacturing process, external environments, and deployment of TOFs [9].

The geometric asymmetry of TOFs breaks modal degeneracy, resulting in birefringence [24,25] and frequency splitting of mechanical modes [22]. The geometric ellipticity of TOFs is a critical factor in causing nondegenerate modes. however, it has not yet received sufficient attention. On the one hand, TOF geometric ellipticity profoundly influences both optical and mechanical performance [24–26], including the stability of mechanical modes, the integrity of optical modes, and coupling efficiency with other optical devices [27]. Any variance in ellipticity can significantly alter TOF birefringence and device interoperability [28, 29]. On the other hand, the integration of micro/nanostructure resonators with TOFs results in the formation of various devices with enhanced detection sensitivity [30]. If the TOF integrated in these devices has geometric ellipticity, which can support a vectorial characteristic to the vibrational modes, the degrees of freedom for optical modulation of the TOF can be expanded. This presents a richer palette of possibilities for application in various domains.

Nondestructive measurements and high-precision control of TOF geometric ellipticity are of paramount importance. Experimentally, the cross-sectional shape of the TOF becomes elliptical due to temperature variations and positional discrepancies between the hydrogen–oxygen flame and the TOF during the fabrication process [29]. Various schemes for measuring TOF geometric ellipticity, such as optical microscopy imaging and scanning electron microscopy (SEM), have been proposed [29, 31]. However, the optical diffraction limit poses a significant barrier to accurate characterization at the subwavelength scale. While SEM offers nanoscale resolution, the requirement for a conductive coating during SEM inspections inadvertently leads to the degradation of TOFs, compromising their suitability for subsequent utilization [32]. In recent years, various nondestructive techniques to measure the diameter of TOFs have emerged [33–36], including techniques based on external gratings [37], optical diffraction patterns [38], optical scattering phenomena [39–41], and the exploitation of optical harmonics as well as mode interference [42–45]. However, the accuracy of these methodologies has not yet reached the level of determining the TOF cross-sectional ellipticity nondestructively.

In this paper, we present a powerful method for measuring the nondegenerate IFMMs of TOFs with high aspect ratios by utilizing near-field scattering, and the TOF geometric ellipticity can be determined. Through Finite Element Method (FEM) simulations, the relationship between the two nondegenerate IFMMs frequency ratio of simplified TOFs and their ellipticities is derived. The simulation results have been successfully confirmed by experiments based on the near-field scattering of a TOF caused by a hemispherical microfiber tip (MFT). This method has the ability to accurately measure the ellipticity of optical waveguides. This research not only advances the fundamental understanding of the mechanical properties of optical waveguides but also heralds a significant stride forward in the development and calibration of motion sensors characterized by high sensitivity and broad responsiveness. This approach expands the applicability of IFMM measurements to a broad range of micro/nanophotonic structures fabricated from various materials or complex photonic structures, harnessing the potential of non-degenerate mechanical modes for advanced applications [46–48].

## 2. Models and simulations

TOFs can be simplified as suspended optical nanowaveguides, and the intrinsic mechanical mode amplitudes depend on the TOF profile. The intrinsic mechanical modes arise from atomic Brownian motion within the material and can be categorized into four types: flexural, longitudinal, torsional and breathing modes. The flexural modes can be detected without the need for external excitation, reflecting their intrinsic vibrational characteristics.Among these modes, IFMMs exhibit a remarkably broad bandwidth and are densely distributed in the frequency range from kilohertz to megahertz [22].

The TOF's IFMMs are concentrated in the waist region because a larger TOF radius will cause a smaller vibration amplitude [22]. For the convenience of numerical simulation via the FEM, we employ a simplified TOF with a uniform radius and elliptical cross section to simulate the nondegenerate modes induced by its ellipticity. An illustration of the model is shown schematically in Figure 1 (a). According to the actual experiments, the waist radius $r$ is set to approximately 250 nm, and the length is set to $L = 5\ mm$. Thus, the TOF exhibits an aspect ratio $R_a = L/r$ as high as 10000. The material of the TOF is silica (refractive index $n = 1.45$) [39]. The inset of Fig. 1 (a) illustrates the TOF's elliptical cross section with radii $r_1$ and $r_2$ along the two orthogonal principal axes, called the major and minor axes. The IFMM eigenfrequencies of the TOF with a uniform radius are given by [49]

$$f^n = \sqrt{\frac{E}{\rho} \frac{(2n+1)^2 \pi r}{16 L^2}}, \quad (1)$$

where $n$ is the IFMM order. The Young's modulus is $E = 73\ GPa$, and the mass density is $\rho = 2320\ kg/m$. A TOF with an elliptical cross section leads to two unique nondegenerate IFMMs. The two orthogonal modes vibrate along the major or minor axis with splitting vibration frequencies. Figure 1 (b) and (c) show typical IFMMs along the major and minor axes, respectively. The mode order is four. The origin of the coordinate axis is at the center of the TOF.

The ratio of two nondegenerate IFMMs frequencies (detailed in Supplemental Material Appendix A) is expressed by

$$\eta_f^n = \frac{\kappa(n)}{\varepsilon}, \quad (2)$$

where $\eta_f^n = f_{major}^n / f_{minor}^n$ represents the frequency ratios for the n-order IFMMs and where $f_{major}^n$ and $f_{minor}^n$ represent the frequencies of the n-order IFMMs along the major and minor axes, respectively. $\varepsilon = r_1/r_2$ is the TOF ellipticity. Empirical models capturing the convergence factors, denoted as $\kappa(n) = Ce^{-\alpha n}$, for $n^{th}$ order IFMMs can be deduced from experimental observations and computational simulations. Here, $C$ and $\alpha$ serve as fitting parameters that are empirically determined to best describe the convergence behavior.

Figure 2 (a) and (b) show the vibrational amplitude as a function of the TOF axis (z axis) for the $1^{st}$, $7^{th}$ and $15^{th}$ IFMMs with $\varepsilon = 1.0000$, $R_a = 6000$ and $\varepsilon = 1.0060$, $R_a = 10000$, respectively. TOFs exhibit larger mechanical wavelengths at the first order mode (order number 1), resulting in an asymmetric distribution of the vibration amplitude along the TOF axis. This asymmetry arises from the interaction between two degenerate modes, leading to mode hardening or softening [46]. This effect is particularly pronounced in TOFs with high aspect ratios. For higher-order modes, the amplitude distribution tends to be consistent for two orthogonal modes along the x- and y-axes. Figure 2 (c) shows the inverse of the frequency ratio for the IFMMs as a function of the mode order for the TOF with different ellipticities. The major radius $r_1$ is 250.0 nm, and the minor radii $r_2$ is 247.5, 248, 248.5, 249, 249.5 and 250.0 nm, corresponding to ellipticities of 1.0101 (green curve), 1.0081 (yellow curve), 1.0060 (blue curve), 1.0040 (red curve), 1.0020 (purple curve) and 1.0000 (black curve), respectively. For a lower order than 10, the inverse of the frequency ratio for the IFMMs has nonlinear convergence with respect to the mode order, and the inverse of the frequency ratio for the IFMMs is greater than one over the ellipticity, as shown in Figure 2 (c). The nonlinearity results from frequency hardening or softening caused by the high aspect ratio of the TOF [46]. Nevertheless, the inverse of the frequency ratio for IFMMs tends toward a constant with increasing mode order. The constants are equal to the TOF ellipticity. The disparity in the two nondegenerate IFMMs, referred to as frequency splitting, becomes increasingly distinct as the ellipticity of the elliptical TOF escalates (detailed in Supplemental Material Appendix B). Figure 2 (d) illustrates the relationship between the inverse of the frequency ratio for the IFMMs and the one over the ellipticity (orange solid dots), and the orange solid line depicts a linear fit. The major radius is $r_1 = 250.0$ nm , and the mode order is 10. Thus, TOF ellipticity can be determined by the frequency ratio of two higher-order nondegenerate IFMMs.

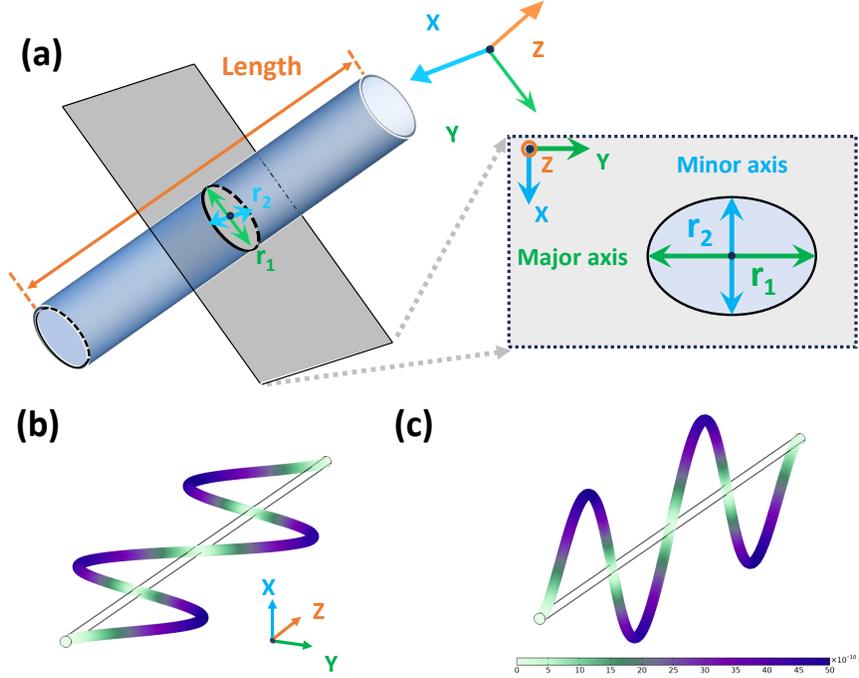

Fig. 1. (a) Simplified TOF with an elliptical cross section. Inset: TOF elliptical cross section with radii $r_1$ and $r_2$ along the major and minor axes. (b) and (c) are the $4^{th}$ IFMMs along the major and minor axes, respectively.

In addition, to eliminate the influence of the TOF aspect on the frequency ratio and to avoid measurement errors in the TOF ellipticity, we simulate the relationship between the aspect ratio and the frequency ratios of the eigenfrequencies for various aspect ratios. The results detailed in the Supplemental material Appendix C show that the aspect ratio has no effect on the frequency ratio for higher-order modes. Thus, the higher-order modes are chosen to measure the frequency ratio of nondegenerate modes and determine the TOF ellipticity.

### 3. Experiments

A TOF-MFT system based on near-field scattering is used to measure the frequency ratio of two nondegenerate modes and to determine the TOF ellipticity . TOFs with subwavelength diameters exhibit a significant evanescent field on the surface. When a hemispherical MFT is immersed into the evanescent field on the TOF surface, it leads to substantial scattering loss in the light transmitted through the TOF [50]. The MFT enables precise positioning within the evanescent field of the TOF, with sub-nanometer control over the distance between the MFT and the TOF. Due to its much larger size compared to the TOF, the MFT's alignment perpendicular to the TOF axis is relatively insensitive. Furthermore, When the diameter of the TOF is small, the presence of the MFT not only alters the effective refractive index outside the fiber but also extends the penetration depth of the evanescent fields. The TOF transmission is susceptible to oscillation in the TOF-MFT gap fluctuations caused by the IFMMs of the TOFs, allowing highly sensitive detection of the TOF's IFMMs on the basis of the spectrum analysis of the TOF transmission. The MFT and TOF are fabricated via a $CO_2$ laser and flame-brushing technique, respectively. Detailed descriptions of the fabrication setups and procedures for both TOFs and MFTs can be found in the references [39, 50].

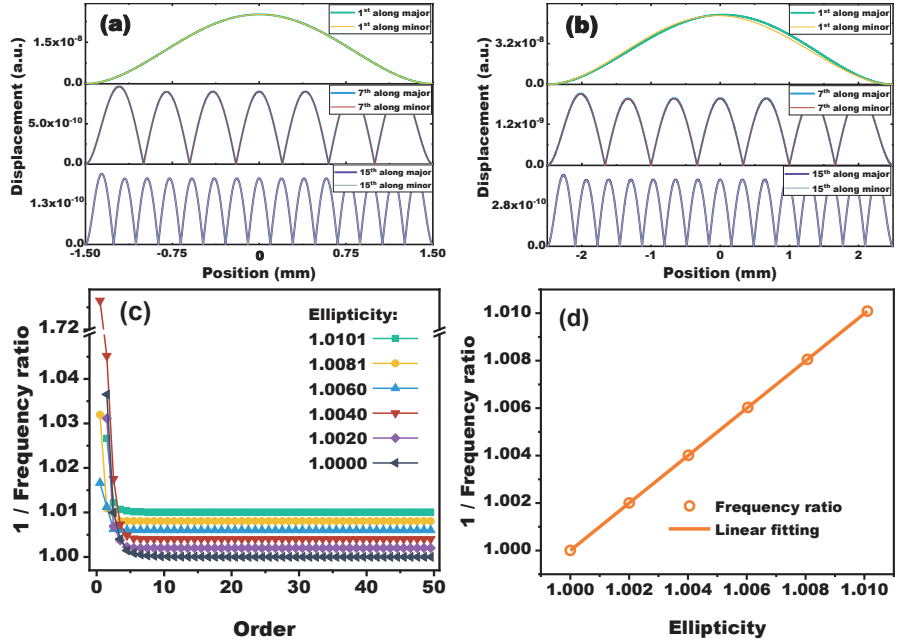

Fig. 2. (a) and (b) show the vibrational amplitude as a function of the TOF axis (z-axis) for the $1^{st}$, $7^{th}$ and $15^{th}$ IFMMs with $\varepsilon = 1.0000$, $R_a = 6000$ and $\varepsilon = 1.0060$, $R_a = 10000$, respectively. (c) The inverse of the frequency ratio for IFMMs as a function of the mode order for the TOF with different ellipticities. The ellipticities are 1.0101 (green curve), 1.0081 (yellow curve), 1.0060 (blue curve), 1.0040 (red curve), 1.0020 (purple curve) and 1.0000 (black curve). (d) The relationship between the inverse of the frequency ratio and the one over the ellipticity (orange solid dots) and the orange solid line depicts a linear fit.

Figure 3 illustrates the TOF-MFT system employed for measuring the flexural modes of the TOF [22]. The TOF and the MFT are affixed to a piezoelectric translation stage (ANPx51/ANPz51, attocube), enabling fine adjustments along the x, y and z axes to ensure subnanometer precision when controlling the distance between the MFT and the TOF. The entire system is enclosed within a high vacuum chamber with pressures as low as $10^{-4}$ Pa to avoid the influence of air. A laser with a wavelength of 852 nm (TLB-6716-P, New Focus) is coupled into a 50:50 fiber beam splitter. One output of the splitter is connected to the TOF as a probing light, whereas the other is coupled to a fiber variable attenuator (VOA850-APC, Thorlabs, not shown in Figure 3) to serve as a reference light. Throughout the experiment, the probe light power was maintained at 20 $\mu W$ to reduce the heating of the TOF and to ensure a satisfactory signal-to-noise ratio. The two beams are detected by a balanced photodetector (PDB420A, Thorlabs) for balanced amplification. The detector's output is then fed into a spectrum analyzer or an oscilloscope. Figure 3 (b) shows a schematic of the coupling mechanism between the MFT and the TOF. The angle between the MFT axis and the TOF major axis is denoted as $\theta$, as shown in Figure 3 (b). The gap between the TOF and the MFT is approximately 37 nm. The detection sensitivity reaches the highest level along the MFT axis because the gap size between the MFT and the TOF is modulated to the maximum when the TOF vibrates along the MFT axis. The detection sensitivity is lowest when the TOF vibrates along the MFT radial direction because the gap modulation amplitude of the MFT radial direction is $10^{-3}$ of the MFT axis direction for the same TOF vibration amplitude. Thus, the TOF-MFT system is a powerful tool for performing vector measurements of TOF mechanical vibration [22].

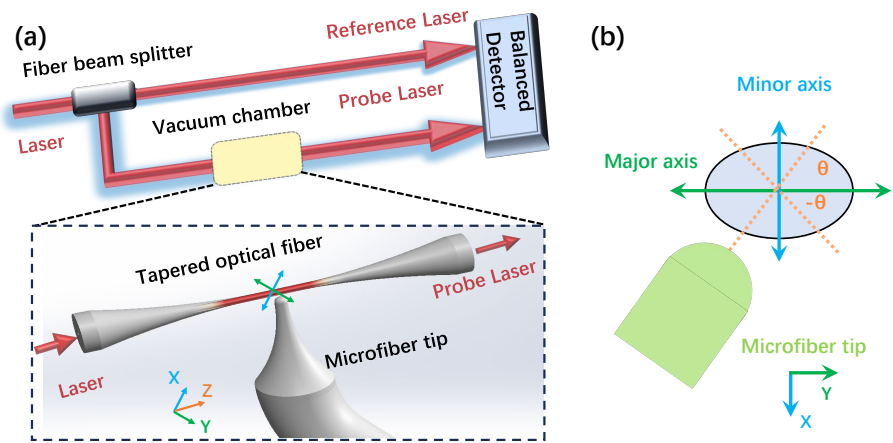

Fig. 3. (a) Experimental setup for detecting the nondegenerate intrinsic flexural mechanical modes. (b) Schematic of the coupling between an MFT and a TOF. The two orthogonal nondegenerate IFMMs vibrating along the TOF minor axis and the major axis. The angle between the MFT axis and the TOF major axis is denoted as $\theta$.

The typical spectrum of the nondegenerate IFMMs is illustrated in Figure 4 (a), and the mode order is 143. The blue circles represent the experimental data, whereas the red, green and blue curves represent Lorentzian fits. Two eigenfrequencies of nondegenerate IFMMs can be distinguished from the spectrum. The double peaks result from the two orthogonal nondegenerate IFMMs vibrating along the TOF minor axis and the major axis. Both nondegenerate IFMMs have projection components of vibration amplitude onto the MFT axis direction. The inset in Figure 4 (a) shows the full frequency–range spectrum (detailed in Supplemental Material Appendix D). In the low-frequency region, where background noise is higher, a notable fluctuation in the amplitude can be observed. This is due to environmental mechanical vibrations amplifying the intrinsic modes of the TOF, leading to increased sensitivity and observable variations. The total number of IFMMs detected reaches 267, and the maximum frequency reaches 1.2 MHz. Double peaks can be observed for any order of IFMMs. Figure 4 (b) is plotted to illustrate the inverse of the frequency ratio (red open circles) for the IFMMs and frequency splitting (green solid squares) of all the nondegenerate IFMMs as a function of the mode order. The red and green curves indicate the linear and exponential fits, respectively. Experimentally, the inverse of the frequency ratio for the nonlinear convergence process of these modes of IFMMs is also characterized. The inverse of the frequency ratio for IFMMs decays with increasing order when the order is lower than 24, whereas the inverse of the frequency ratio for IFMMs tends toward a constant for IFMMs with mode orders higher than 24. This trend is the same as the simulation results depicted in Figure 2 (a). The experimentally measured exponential decay rates in Figure 4(b) do not fully match the simulations shown in Figure 2(c), which could be due to the specific boundary conditions and geometry of the TOF, imperfections in the TOF fabrication process and the presence of axial stresses. The measurement results show that the frequency splitting reaches as high as 731 Hz and that the inverse of the frequency ratio for the IFMMs reaches 1.0014 when the order number is 155 and the frequency is 503 kHz. As the mode order increases and is greater than 24, the frequency splitting increases linearly. The larger frequency splitting for higher orders makes it possible to measure smaller frequency changes. This can increase the measurement accuracy of the frequency ratio and TOF ellipticity calibration, which play pivotal roles in enhancing the accuracy of vector measurements.

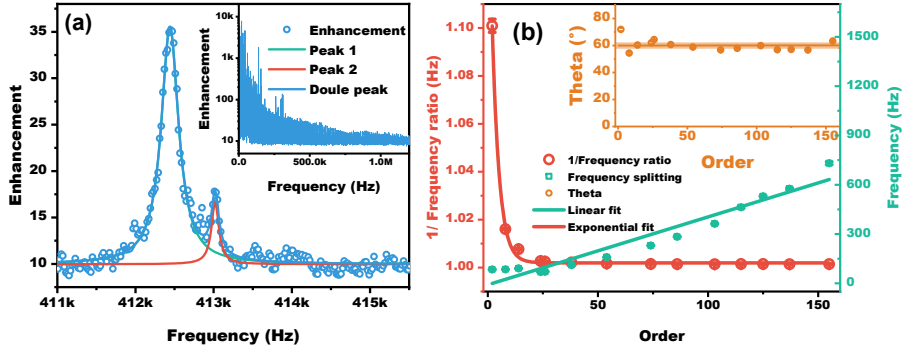

Fig. 4. (a) Typical frequencies of the high-order nondegenerate mode-splitting flexural modes. Inset: full-range enhancement spectrum. (b) The inverse of the frequency ratio for nondegenerate IFMMs and frequency splitting as a function of IFMM order. Inset: angle between the principal axes and the 'TOF-MFT' plane of the IFMMs as a function of frequency.

By employing the frequency ratios of the higher-order IFMMs, the TOF ellipticity can be determined. For the TOF's elliptical cross section, there are two principal area moments of inertia, which correspond to the two orthogonal axes. The area moments of inertia along the minor and major axes are $I_{minor} = \int_A y^2 dA = \pi r_1^3 r_2/4$ and $I_{major} = \int_A y^2 dA = \pi r_1 r_2^3/4$ [46], respectively. The two nondegenerate IFMMs share an identical modal mass and damping ratio, and they significantly differ in their modal stiffness and area moment of inertia. According to the equipartition theorem [46], the two modes have different thermally driven power spectral densities, and the mean squared deflection $\langle d_1^2 \rangle$ and $\langle d_2^2 \rangle$ of the $n^{th}$ eigenmode along the major or minor axis are expected to scale with the square of the flexural frequency [51]:

$$\frac{\langle d_2^2 \rangle}{\langle d_1^2 \rangle} = \frac{f_{major}^2}{f_{minor}^2}, \tag{3}$$

where the bracket $\langle \rangle$ denotes the time average; however, there is a nonright angle between the MFT axis direction and the TOF major axis. The measured mean squared deflections $\langle d_{1,meas.}^2 \rangle$ and $\langle d_{2,meas.}^2 \rangle$ have projection components of the vibration amplitude onto the MFT axis direction. The projection components depend on the angle $\theta$ between the MFT axis and the major axis. The relationship between the measured mean squared deflection and the mean

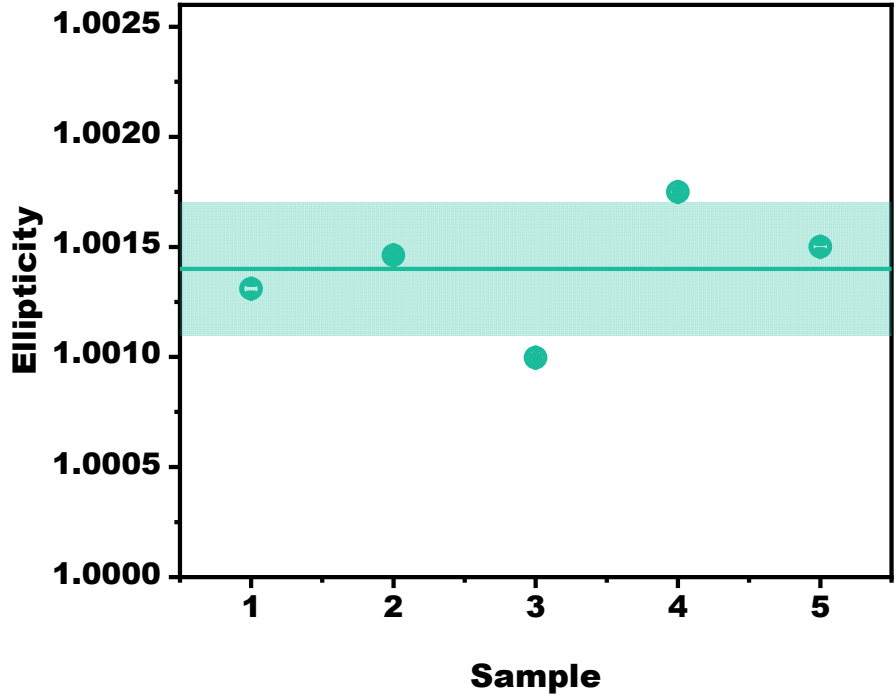

Fig. 5. Ellipticity of five TOF samples. The green open circles are the experimental results, whereas the green dashed line and green shaded area indicate the average and deviation for the five TOF samples.

squared deflection along the major or minor axis is as follows:

$$\langle d_{1,\text{meas.}}^2 \rangle = \langle d_1^2 \rangle \sin^2 \theta, \tag{4}$$

$$\langle d_{2,\text{meas.}}^2 \rangle = \langle d_2^2 \rangle \cos^2 \theta. \tag{5}$$

Equations (4) and (5) can be solved to yield the angle of the measurement

$$\theta = \arctan \sqrt{\frac{\langle d_{1,\text{meas.}}^2 \rangle}{\langle d_{2,\text{meas.}}^2 \rangle} \cdot \frac{f_{\text{major}}^2}{f_{\text{minor}}^2}}. \tag{6}$$

The angle $\theta$ (orange open circles) between the principal axes as a function of mode order is shown in the inset of Fig. 4 (b). The blue curves are exponential fits, and the orange line and orange shaded area indicate the average and deviation, respectively, for the angle $\theta$. The angle remains constant at 60 ± 4 with increasing order number. The cross-section ellipticity of the TOF can be determined through the equation (2).

The ellipticities of the five TOF samples are shown in Figure 5. The green open circles indicate the experimental results, with the error bars on each data point representing the measurement uncertainties of the five samples, reflecting the precision of the experimental measurements (approximately 1.4E-5 ). The green line and the green shaded area represent the average value (1.0014) and standard deviation ( 3E-4), respectively. All samples are fabricated using the same setup and procedure. It is noteworthy that the measurement uncertainties are substantially smaller than the variations introduced by the fabrication process. The average TOF radius is approximately 260 ± 5 nm calibrated according to reference [39]. The average ellipticity for 5 samples is 1.0014 ± 0.0003. Notably, the measurement deviation of the TOF ellipticity obtained through this scheme can reach 0.16 nm despite the measurement uncertainty of approximately 5 nanometers for the TOF radius.

## 4. Conclusions

In summary, we simulate numerically and implement a method to measure the nondegenerate IFMMs and to quantify the TOF ellipticity on the basis of the nondegenerate modes successfully. The simulations are implemented and show that the inverse of the frequency ratio for the IFMMs has nonlinear convergence with respect to the mode order when the order number is smaller than 10, whereas the inverse of the frequency ratio for the IFMMs tends toward a constant with higher-order modes. The experimental results agree well with the simulation results. Typically, the frequency splitting reaches as high as 731 Hz, and the inverse of the frequency ratio for IFMMs reaches 1.0014 when the order number is 155 and the frequency is 503 kHz. According to the frequency splitting of the higher-order modes, the TOF geometric ellipticity can be determined with a subnanometer-level accuracy of 0.16 nm. Furthermore, the larger frequency splitting of its higher-order mechanical vibration modes offers enhanced and high resolution. There are several methods to further enhance measurement performance: 1) The distance between MFT-TOF also affects the sensitivity of the measurement results, thus, the smaller distance which causes the greater scattering loss leads to higher signal-to-noise ratio, better sensitivity and a wider measurable frequency range. 2) Enhancing vibration isolation is vital to improve measurement accuracy, particularly for lower-order modes. By isolating low-frequency disturbances, the overall precision and reliability of this method can be significantly improved. 3) TOFs with MFT excel in detecting subtle changes in electric and magnetic fields and microscopic displacements due to their minuscule mass and exceptional optical scattering sensitivity. 4) The method can be improved to measure other types of modes, for example, putting two or more tips in different directions of TOFs. By leveraging these traits, these methods can be fine-tuned to achieve unparalleled sensitivity and measurement precision.

This method is pivotal for understanding the mechanical behavior of optical waveguides and represents a novel advancement in the research of micro/nanomechanics. This method allows the measurable range to be broadened even further [50]. The low elliptical cross-section with a low area moment difference ensures nearly identical sensitivities of the system to environmental physical changes along the major or minor axis. Exploiting the subtle variations in ellipticity provides additional degrees of freedom for measurement process. On the basis of the high-sensitivity enhanced measurements of optical waveguide ellipticity and the mechanical vibration direction, this approach offers significant advantages in vectorial property characterization. The subnanometer measurement accuracy makes it suitable for characterizing the microstructure morphology and vectorial properties of micro/nanoscale devices such as cantilevers [46,52]. This geometric asymmetry of TOFs may be useful for understanding the behavior of laser-tweezered atoms when flexural vibrations can create mechanical heating [53,54]. Optical waveguides with both elliptical cross-sections, when combined with solid-state emitters, can be utilized for vectorial measurements of physical quantities such as vibrations, magnetic fields, and electric fields. The nondegeneracy of TOF mechanical modes enables broad application prospects in optomechanics [49,55], microdisplacement measurements [56–59], micro-Newton force sensing [60–62], and multidimensional quantum states [63], among other areas [64–70]. By leveraging the unique properties of nondegenerate mechanical modes, this work contributes to the development of next-generation photonic and quantum devices and expanding their potential applications.

**Funding.** Innovation Program for Quantum Science and Technology (2023ZD0300400); National Key Research and Development Program of China(2021YFA1402002); National Natural Science Foundation of

# Supplemental Material for Sub-nanometer measuring ellipticity of a suspended optical nanowaveguides based on nondegenerate mechanical modes

CHENXI WANG[1], LIJUN SONG[1], JIANTING WANG[1], JING ZHOU[1], KANGJIE FENG[1], QIANG ZHANG[1,2], CHANG LING ZOU[3,1], GANG LI[1,2], PENGFEI ZHANG[1,2,*], AND TIANCAI ZHANG[1,2,**]

[1]*State Key Laboratory of Quantum Optics and Quantum Optics Devices, Institute of Opto-electronics, Shanxi University, Taiyuan, Shanxi 030006, People's Republic of China*
[2]*Collaborative Innovation Center of Extreme Optics, Shanxi University, Taiyuan, Shanxi 030006, People's Republic of China*
[3]*Key Laboratory of Quantum Information, Chinese Academy of Sciences, University of Science and Technology of China, Hefei 230026, People's Republic of China*
*\* zhangpengfei@sxu.edu.cn*
*\*\* tczhang@sxu.edu.cn*

## 1. Appendix A

To derive the ratio of two nondegenerate IFMMs frequencies, the area moments of inertia for an elliptical TOF cross-section are considered. The area moment of inertia (also known as the second moment of area) is a measure of how the area of a shape is distributed relative to an axis. For an ellipse TOF, the area moments of inertia along the major axis and the minor axis can be calculated starting with the general equation for an ellipse:

$$\frac{x^2}{r_1^2} + \frac{y^2}{r_2^2} = 1, \tag{1}$$

where $r_1$ and $r_2$ are radii along the two orthogonal axes, called the major and minor axes, respectively.

For the major axis, the area moment of inertia is given by

$$I_{major} = \int_{-r}^{r} \int_{-\sqrt{r_2^2(1-x^2/r_1^2)}}^{\sqrt{r_2^2(1-x^2/r_1^2)}} y^2 \, dy \, dx = \int_{-r}^{r} \left[\frac{1}{3} y^3\right]\int_{-\sqrt{r_2^2(1-x^2/r_1^2)}}^{\sqrt{r_2^2(1-x^2/r_1^2)}} dx = \int_{-r}^{r} \frac{2}{3} r_2^2 (1 - x^2/r_1^2)^{3/2} dx. \tag{2}$$

By substituting $u = x/r_1$ and $du = dx/r_1$, we can obtain

$$I_{major} = \frac{2}{3} r_2^2 r_1 \int_{-1}^{1} \left(1 - u^2\right)^{3/2} du. \tag{3}$$

The integral can be simplified to:

$$I_{major} = \frac{\pi}{4} r_1 r_2^3. \tag{4}$$

For the minor axis, this integral can be simplified to:

$$I_{minor} = \frac{\pi}{4} r_1^3 r_2. \tag{5}$$

For a TOF described by Euler–Bernoulli beam theory, the linear resonance frequency of each flexural mode is proportional to the square root of the bending stiffness related to motion in the direction of the mode

$$f_n^j = \frac{\omega_n^j}{2\pi} = \frac{\kappa(n)}{2\pi L^2}\sqrt{\frac{EI^j}{m}} \propto \sqrt{D_j}. \tag{6}$$

where $D_j$ is the bending stiffness with E as the Young's modulus and I is the second moment of the area associated with the bending of the TOF in the direction of mode j (major or minor). L is the length of the TOF, m is the mass per unit length, and $\kappa(n)$ is a mode-dependent coefficient.

The ratio of two nondegenerate IFMMs as a function of the ellipticity of the elliptical TOF is

$$\eta_f^n = \frac{f_{major}}{f_{minor}} = \kappa(n)\sqrt{\frac{I_{major}}{I_{minor}}} = \kappa(n)\sqrt{\frac{r_1 r_2^3}{r_1^3 r_2}} = \kappa(n)\frac{r_2}{r_1} = \frac{\kappa(n)}{\varepsilon}. \tag{7}$$

## 2. Appendix B

Figure S 1 depicts the $23^{rd}$ and $24^{th}$ IFFMs eigenfrequencies along the major and minor axes, respectively. The vibration frequencies are correlated with the TOF radius. The eigenfrequency along the major axis is largely unaffected by ellipticity, whereas the eigenfrequency along the minor axis decreases as the ellipticity increases. The disparity in the vibrational frequencies, referred to as frequency splitting, becomes increasingly distinct as the ellipticity of the TOF escalates. The inset of Figure S 1 shows the frequency splitting ($f_{major} - f_{minor}$) of the IFMMs as a function of the mode order. This phenomenon highlights the sensitivity of the vibrational characteristics to the geometric asymmetry introduced by the elliptical shape. Figure S 1 $23^{rd}$ and $24^{th}$ IFFMs eigenfrequencies along the major and minor axes, respectively. The inset shows the TOF frequency splitting of two nondegenerate IFMMs as a function of the mode order.

## 3. Appendix C

Nondegenerate IFMMs frequency ratio as a function of order with various TOF aspect ratios. To verify that the aspect ratio $R_a$ of an elliptical TOF does not affect the eigenfrequency splitting of higher-order modes, we simulate the modes of the elliptical TOF with various aspect ratios, as shown in Figure S 2. The green, yellow, blue, red, purple, and black curves represent the relationships between the eigenfrequency ratios of the two nondegenerate modes and the mode order for TOFs with aspect ratios $R_a$ of 2000, 6000, 10000, 14000, 18000 and 22000, respectively. The TOF radii are $r_1 = r_2 = 250$ nm, while the TOF lengths are 1, 3, 5, 7, 9, and 11 mm. The inset in Figure S 2 provides a magnified inverse of the frequency ratio for IFMMs with mode orders ranging from 10–17. For higher-order modes, the inverse of the frequency ratio for IFMMs is equal to 1 for different TOF aspect ratios and does not depend on the mode order or the TOF aspect ratio.

Figure S 2 Relationships between the eigenfrequency ratios of the two nondegenerate modes and mode order for the TOF with aspect ratios $R_a$ of 2000 (green), 6000 (yellow), 10000 (blue), 14000 (red), 18000 (purple) and 22000 (black). Inset: magnified nondegenerate IFMMs eigenfrequency ratios with mode orders ranging from 10–17.

## 4. Appendix D

The enhancement spectrum is the ratio of the enhanced spectrum to the spectral background without the MFT. As shown in Fig. S3, the blue curve is the enhancement spectrum with a 37 nm gap between the TOF and the MFT.

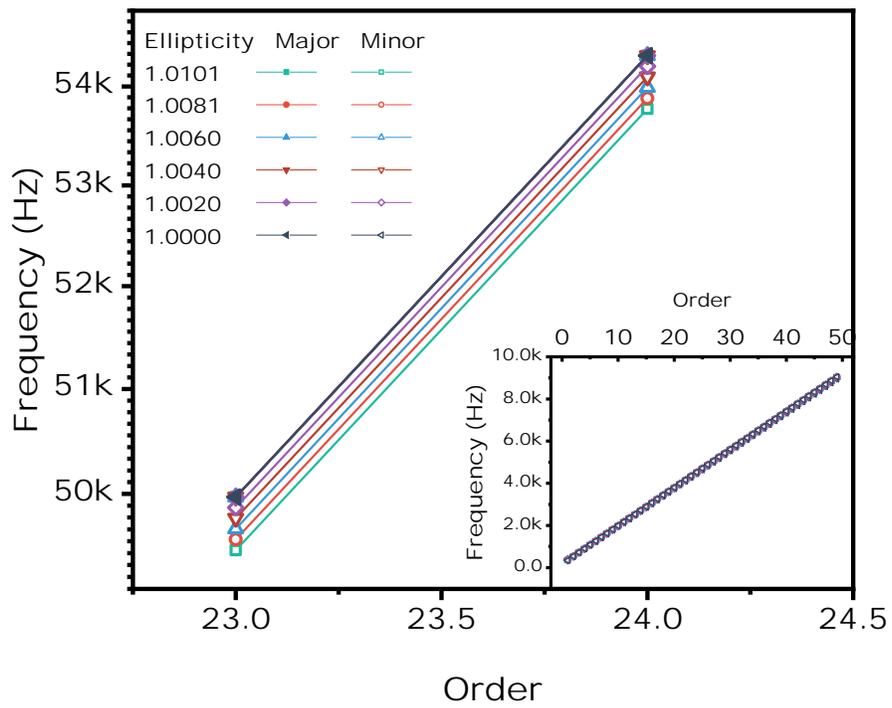

Fig. S 1. $23^{rd}$ and $24^{th}$ IFFMs eigenfrequencies along the major and minor axes, respectively. The inset shows the TOF frequency splitting of two nondegenerate IFMMs as a function of the mode order.

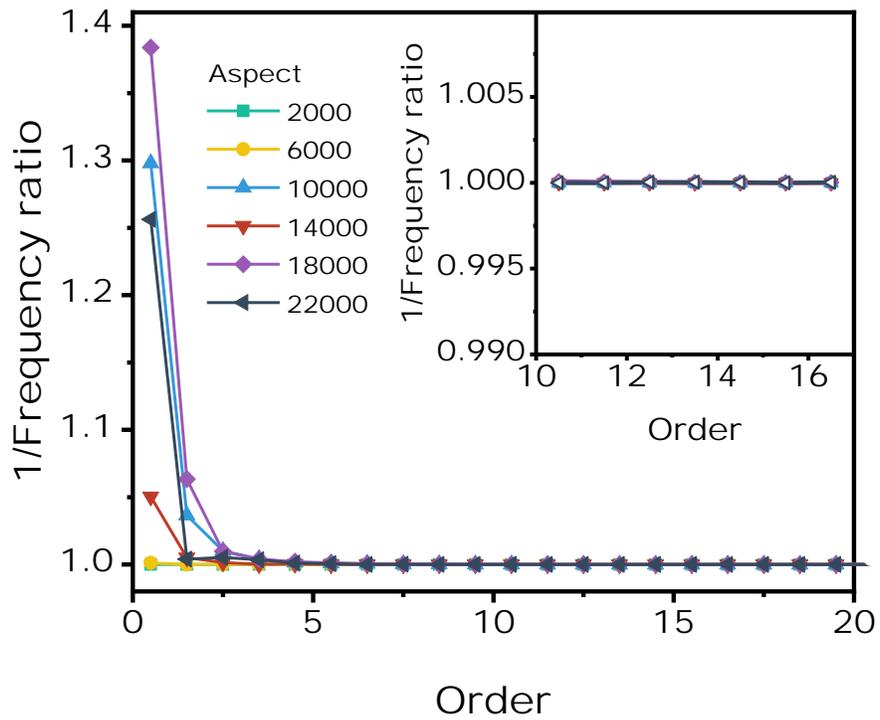

Fig. S 2. Relationships between the eigenfrequency ratios $R_a$ of the two nondegenerate modes and mode order for the TOF with aspect ratios of 2000 (green), 6000 (yellow), 10000 (blue), 14000 (red), 18000 (purple) and 22000 (black). Inset: magnified nondegenerate IFMMs eigenfrequency ratios with mode orders ranging from 10–17.

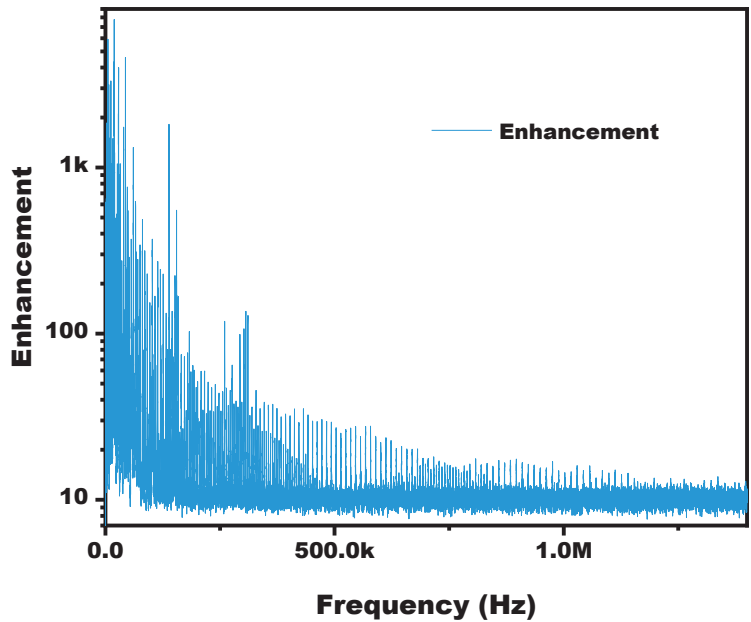

Fig. S 3. Full frequency–range enhancement spectrum of TOF's nondegenerate IFMMs.